\documentclass[twocolumn,floatfix,aps,eqsecnum,superscriptaddress,prb]{revtex4-2}
\usepackage{graphicx}
\usepackage{amsmath}
\usepackage{amssymb}
\usepackage{amsthm}
\usepackage{dsfont}
\usepackage{bm}
\usepackage{physics}
\usepackage{hyperref}
\usepackage{float}
\usepackage{xcolor}
\usepackage{soul}
\newtheorem*{remark}{Proposition}
\newcommand\varpm{\mathbin{\vcenter{\hbox{%
  \oalign{\hfil$\scriptstyle+$\hfil\cr
          \noalign{\kern-.3ex}
          $\scriptscriptstyle({-})$\cr}%
}}}}

\begin{document}

\title{Dynamical simulation of the injection of vortices into a Majorana edge mode}

\author{I. M. Fl\'or}
\thanks{Now at Department of Physics, KTH Royal Institute of Technology, Stockholm 106 91, Sweden}
\affiliation{Instituut-Lorentz, Universiteit Leiden, P.O. Box 9506, 2300 RA Leiden, The Netherlands}

\author{A. Don\'is Vela}
\affiliation{Instituut-Lorentz, Universiteit Leiden, P.O. Box 9506, 2300 RA Leiden, The Netherlands}
\author{C. W. J. Beenakker}
\affiliation{Instituut-Lorentz, Universiteit Leiden, P.O. Box 9506, 2300 RA Leiden, The Netherlands}
\author{G. Lemut }
\thanks{Now at Dahlem Center for Complex Quantum Systems and Fachbereich Physik, Freie Universit\"at Berlin, 14195 Berlin, Germany}
\affiliation{Instituut-Lorentz, Universiteit Leiden, P.O. Box 9506, 2300 RA Leiden, The Netherlands}


\date{July 2023}
\begin{abstract} 
The chiral edge modes of a topological superconductor can transport fermionic quasiparticles, with Abelian exchange statistics, but they can also transport non-Abelian anyons: Edge-vortices bound to a $\pi$-phase domain wall that propagates along the boundary. A pair of such edge-vortices is injected by the application of an $h/2e$ flux bias over a Josephson junction. Existing descriptions of the injection process rely on the instantaneous scattering approximation of the adiabatic regime [Beenakker \textit{et al.} Phys.Rev.Lett. {122}, (2019)]
, where the internal dynamics of the Josephson junction is ignored. Here we go beyond that approximation in a time-dependent many-body simulation of the injection process, followed by a braiding of mobile edge-vortices with a pair of immobile Abrikosov vortices in the bulk of the superconductor. Our simulation sheds light on the properties of the Josephson junction needed for a successful implementation of a flying  topological qubit.

\end{abstract}
\maketitle

\section{Introduction}
\label{sec:intro}

A remarkable property of topological superconductors is that two vortices winding around each other exchange a quasiparticle\cite{Rea01,Iva02,Been20}. This ``braiding'' operation is a manifestation of the non-Abelian statistics of the Majorana zero-modes bound to the core of an Abrikosov vortex \cite{Nay03,Das04,Jia14}. 
Because Abrikosov vortices are immobile, typically pinned to defects, winding them is a thought experiment that is not easily implemented \cite{Ma05,Ma06,Vla07}. 

A proposal to mobilize vortices by injecting them into the edge modes of a topological superconductor was suggested by Beenakker \textit{et al}. (Ref.~\onlinecite{Bee08}), where the parity carried by the edge vortices encodes a qubit. After the injection, such edge-vortices can be braided with bulk vortices due to their chiral motion, without requiring any external manipulation. This results in a fermion parity switch (flip of the qubit) between the edges and the bulk that can be detected electrically as an $e/2$ charge pulse when a pair of edge vortices is fused in a normal metal contact \cite{Ada09,Has10}.

\begin{figure}[tb]
\centerline{\includegraphics[width=.93\linewidth]{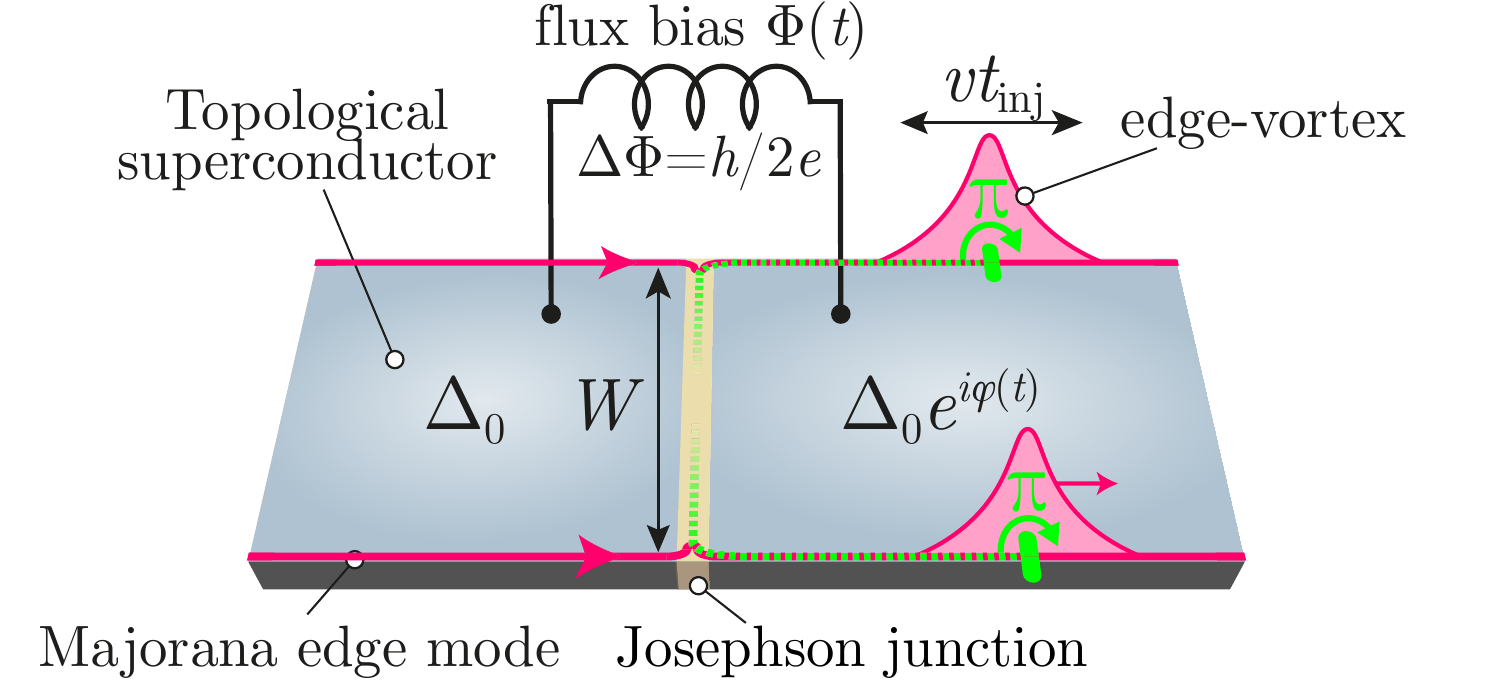}}
\caption{Edge vortex injector \cite{Bee08}, consisting of a Josephson junction in a topological superconductor with co-propagating chiral edge modes. An $h/2e$ flux increment injects a pair of edge-vortices on opposite edges with a protected fermion parity. The corresponding phase domain wall is represented with green lines. The adiabatic description of the injection process assumes that the injection time $t_{\rm inj}=(2\pi\xi_{\rm J}/W)(\mathrm{d}\varphi/\mathrm{d}t)^{-1}$ is long compared to the propagation time $W/v$ along the junction. In this work we relax that assumption, to simulate a device (Fig.~\ref{fig:setup}) where these dynamically injected edge-vortices are braided with Abrikosov bulk vortices. 
}
\label{fig_layout}
\end{figure}

The key component of the braiding device of Ref.~\onlinecite{Bee08} is the edge-vortex injector (see Fig.~\ref{fig_layout}): it consists of a flux-biased Josephson junction, connecting co-propagating chiral edge modes. The application of a flux bias of $h/2e$ increments the superconducting phase $\varphi$ by $2\pi$. For the fermionic edge mode wave functions this amounts to a $\pi$-phase domain wall \cite{Nayak07}, which moves away from the junction with the Fermi velocity $v$, carrying the edge-vortex excitations. The injection process takes a finite time $t_{\rm inj}$, that translates into a finite width $vt_{\rm inj}$ of the domain wall. Given a rate of change $\mathrm{d}\varphi/\mathrm{d}t$, a junction width $W$, and a superconducting coherence length $\xi_{\rm J}$ one has
\begin{equation}
t_{\rm inj}=(2\pi\xi_{\rm J}/W)(\mathrm{d}\varphi/\mathrm{d}t)^{-1}.
\end{equation}
A major simplification of the theoretical description of the injection process arises if $t_{\rm inj}$ is large compared to the propagation time $W/v$, so for a sufficiently slow rate of change $\mathrm{d}\varphi/\mathrm{d}t\ll2\pi v\xi_{\rm J}/W^2$. This is the so-called adiabatic regime, in which one may rely on the instantaneous scattering approximation. Ref.~\onlinecite{Bee08} applies to that regime.  The purpose of the present paper is to relax the adiabatic approximation, to see how large $(v/W)t_{\rm inj}$ should be for the braiding operation to succeed. This is studied via a fully dynamical simulation of the proposed device during the injection, braiding and fusion.

Since an edge vortex is a collective degree of freedom, the dynamics involves the full many-body state. We study it numerically, by means of time-dependent Bogoliubov-de Gennes methods. Our main conclusion is that a factor of two between $t_{\rm inj}$ and $W/v$ is sufficient to avoid the excitations of internal degrees of freedom in the junction that would spoil the fermion parity switch \cite{Sau19,Sch17,JFu21,Sarma09}.

The outline of the paper is as follows: the simulated device and the time-dependent model are introduced in Sec.~\ref{sec:device}. In Sec.~\ref{sec:charge}, we present the results of the braiding protocol which recover the main predictions from the adiabatic theory, namely the charge signature at the exit of the device and the fermion parity exchange of the edges with the bulk. Sec.~\ref{sec:bound_states} describes the excitation dynamics of the junction in the alternative regime $W> vt_\mathrm{inj}$ where the braiding protocol cannot hold. The conclusion is presented in Sec.~\ref{sec:conclusion}.
\section{Model and device}
\label{sec:device}

\subsection{Setup}
We consider the device shown in Fig.~\ref{fig:setup} (a). A quantum anomalous Hall (QAH) insulator ($\mathcal{N}=2$) exhibits an electronic chiral mode (corresponding to two Majorana fermions in the BdG formalism), on each of the two edges \cite{Qi16,Wang15,YWang15}. When the edge of a QAH is proxitimitized by an s-wave superconductor, the fermionic edge mode splits into two spatialy separated co-propagating chiral Majorana fermions, localized at the edges of the superconducting region  \cite{Qi12,Zhang15}. This proximitized system can be described as a topological superconductor ($\mathcal{N}=1$). In our setup, such a topological superconductor (TSC) with two co-propagating Majorana edge modes (Fig.~\ref{fig:setup} (b)) is divided in three sections by two Josephson junctions, each of length $W$ and thickness $w$. The junctions are separated by a distance $L$. Two vortices of flux $\Phi_0=h/2e$ are created in the bulk by an external magnetic field, one of which is in the region between the two junctions. 

{A time-dependent flux bias is applied such that the phase in the middle superconductor is $\varphi(t)$ relative to the others, as in Fig.~\ref{fig_layout}. By increasing the phase $\varphi(t)$ from $0$ to $2\pi$, the effective gap inside the Josephson junctions closes at $\varphi=\pi$ (Fig.~\ref{fig:setup} (c)). In this process, a Josephson vortex \cite{Stern11} passes through each junction, which must locally change the boundary condition from periodic to anti-periodic along the two edges \cite{Nayak07} inducing a phase domain wall in the wave functions over some characteristic time $t_\mathrm{inj}$. This local change of the boundary conditions can be described in terms of an edge vortex field operator $\hat{\mu}(x)$, a collective excitation with non-Abelian statistics \cite{Ada09,Nayak07}. The injected edge-vortices -- one pair at the back junction and another pair at the front junction -- then propagate along the edges with the Fermi velocity $v$. The injection time is given by $t_\mathrm{inj}=(2\pi\xi_\mathrm{J}/W)(\mathrm{d} \varphi(t)/\mathrm{d} t)^{-1}$ where $\xi_\mathrm{J}=\hbar v/\Delta_\mathrm{J}$ {\cite{Bee08}} is the coherence length of the junction. Here $\Delta_\mathrm{J}$ denotes the effective gap in the junction \cite{Fu11} (calculated for an infinite junction as shown in Fig.~\ref{fig:setup} (c)). As long as the characteristic injection time is slow compared to $W/ v$, only the two lowest energy states in the finite junction play a role in the dynamics (see App.~\ref{app:exhaustive}).

The edge-vortices of size $v t_\mathrm{inj}$ then propagate along the edges. The pair of edge-vortices injected at the back overtake a bulk vortex over a distance $L$. This induces a relative sign flip between the edge vortices and effectively results in a quasiparticle being transferred between the edge vortices and the vortices in the bulk. This parity switch of the edge vortices and the bulk vortices is denoted by $P_\mathrm{edges}\rightarrow -P_\mathrm{edges}$ and $P_\mathrm{vortices}\rightarrow -P_\mathrm{vortices}$, i.e. a flip of the qubit encoded in parity of the edge-vortices.

The braiding event can be detected upon the fusion at the exit of the superconductor via a charge measurement. The edge-vortices injected at the front junction produce a charge $e/2$ independently, while the edge-vortices injected at the back junction produce a charge $\pm e/2$ depending on whether they have braided with the bulk vortex. The resulting net charge at the exit is $e(N_\mathrm{vortex}\mod2)$ with $N_\mathrm{vortex}$ the number of vortices in between the two injectors. In Fig.~\ref{fig:intro_snapshots}, the local excitation density and local charge during the braiding protocol are shown for an example simulation.\\

\begin{figure}
    \centering
    \includegraphics[width=\linewidth]{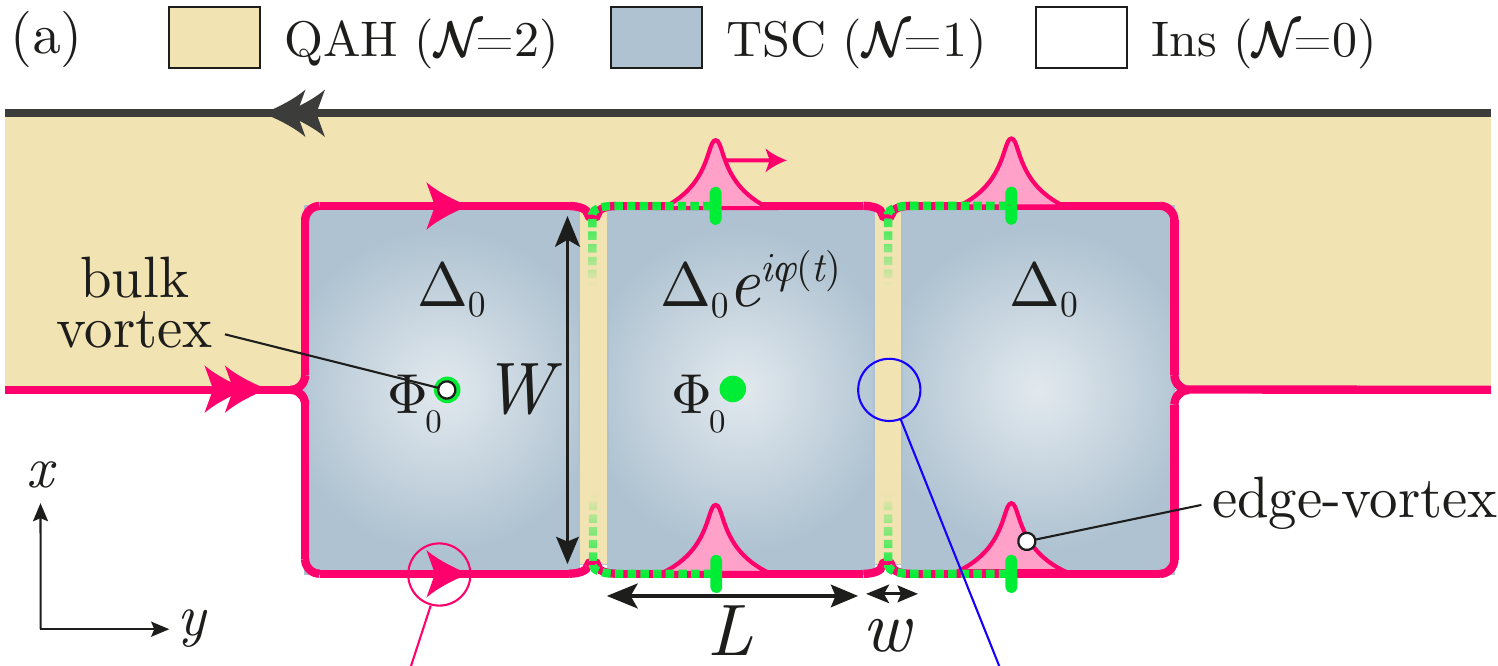}
    \includegraphics[width=.85\linewidth]{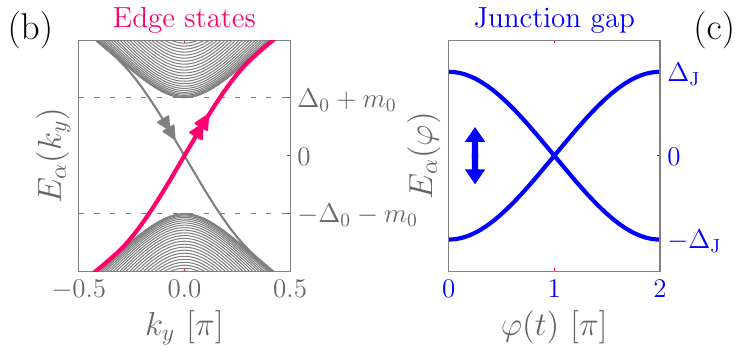}
    \caption{(a) Full braiding device: two injectors (as in Fig.~\ref{fig_layout}) are used to produce pairs of edge-vortices. The pair of edge-vortices at the back exchanges parity with the bulk vortices upon overtaking a bulk vortex, which is detected by an $e/2$ charge measurement at the exit. (b) Dispersion of Majorana edge modes (magenta), calculated for an infinite strip of a topological superconductor ($\mathcal{N}=1$). (c) Lowest energy levels in an infinite Josephson junction (described in Sec. \ref{sec:device}) as a function of the superconducting phase. At $\phi=\pi$ these modes become degenerate and correspond to chiral Majorana edge states propagating along the junction  \cite{Fu11}.}
    \label{fig:setup}
\end{figure}

\begin{figure}
    \centering
    \includegraphics[width=\linewidth]{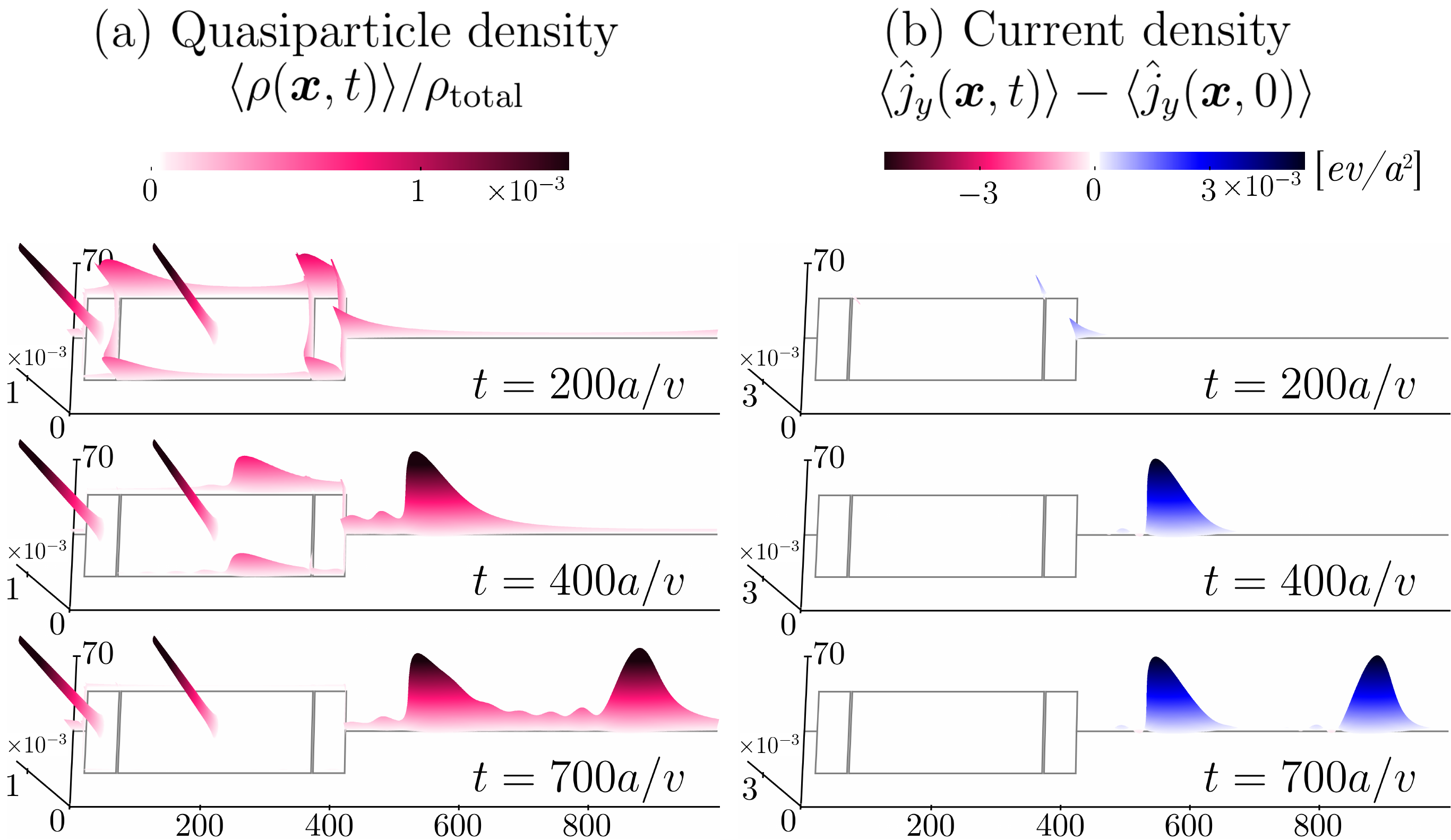}
    \caption{Time snapshots of a dynamical simulation of the full device during the injection and braiding protocol, (a) Bogoliubov quasiparticle density as defined in equation \eqref{eq:local_excitation_density} and (b) current density as defined in \eqref{eq:current_density_at_exit}. In this simulation $vt_\mathrm{inj}=1.5W\ll L$, so the edge-vortices injected at the back and front junction are well separated creating two separate $e/2$ charge pulses upon fusion. An animated version can be found at \cite{supplemental}.
    }
    \label{fig:intro_snapshots}
\end{figure}

\subsection{Hamiltonian}

The device of Fig.~\ref{fig:setup} is simulated using a tight-binding model of a QAH. In the central regions the QAH is proximitized with an s-wave superconductor. The Hamiltonian is given by \cite{Qi12}:
\begin{equation}
    \hat{H}(t) = \frac{1}{2}\sum_{\boldsymbol{x}}\hat{\Psi}^\dagger(\boldsymbol{x})H (\boldsymbol{k},\boldsymbol{x},t)\hat{\Psi}(\boldsymbol{x})
    \label{eq:Hamiltonian}
\end{equation}
where $\hat{\Psi}(\boldsymbol{x}) = (\hat{\psi}_{\uparrow}(\boldsymbol{x}),\hat{\psi}_{\downarrow}(\boldsymbol{x}),\hat{\psi}^\dagger_{\downarrow}(\boldsymbol{x}),-\hat{\psi}^\dagger_{\uparrow}(\boldsymbol{x}))^\intercal$ is the four component Nambu spinor and $H$ is the Bogoliubov-de-Gennes (BdG) Hamiltonian matrix
\begin{equation}
    \begin{split}
    H(\boldsymbol{k}, \boldsymbol{x}, t) = \begin{pmatrix}
        H^e(\boldsymbol{k},\boldsymbol{x})-\mu & \Delta_0 (\boldsymbol{x}) e^{i\vartheta(\boldsymbol{x},t)} \\
        \Delta_0(\boldsymbol{x}) e^{-i\vartheta(\boldsymbol{x},t)}  & \mu-\mathcal{T}H^e(\boldsymbol{k},\boldsymbol{x})\mathcal{T}^{-1}
    \end{pmatrix}
    \end{split}\label{eq:bdg_ham_matrix}
\end{equation}
with $\mu$ the chemical potential and $\mathcal{T}=i\sigma_y\mathcal{K}$ the time-reversal operator ($\sigma_y$ is the second Pauli matrix in the spin degree of freedom and $\mathcal{K}$ denotes complex conjugation). The electronic block is given by:
\begin{equation}
    \begin{split}
        H^e(\boldsymbol{k},\boldsymbol{x}) &= \frac{\hbar v}{a}\left(\sigma_x\sin(k_xa)+\sigma_y\sin(k_ya)\right) \\
        & +(m_0(\boldsymbol{x})+M(\boldsymbol{k}))\sigma_z
    \end{split}
    \label{eq:e_ham}
\end{equation}
where $M(\boldsymbol{k})=\frac{2m_1}{a^2}\left(2-\cos(k_xa)-\cos(k_ya)\right)$ and $\boldsymbol{k}=-i\nabla$. The simulated system is finite in the $x$-direction and anti-periodic in the $y$-direction to ensure that there are no $\boldsymbol{k}=0$ modes in the edges initially \cite{Nayak07,Fu11}. 

The different Chern numbers in the regions of Fig.~\ref{fig:setup} are achieved by different values of $m_0$ and $\Delta_0$:
 \begin{equation}
 \begin{split}
         &m_0(\boldsymbol{x})=-0.5, \ \Delta_0(\boldsymbol{x}) = 0 :\ \boldsymbol{x}\in\mathrm{QAH} \\
         &m_0(\boldsymbol{x})=-0.5, \ \Delta_0(\boldsymbol{x}) = 1:\ \boldsymbol{x}\in\mathrm{TSC}
         \\
         &m_0(\boldsymbol{x})=+\infty, \ \ \Delta_0(\boldsymbol{x}) = 0: \ \boldsymbol{x}\in\mathrm{Ins}
 \end{split}
 \end{equation} %
in units of $\hbar v/a$. The trivial insulating region (Ins) is realized by truncation of the lattice.
Furthermore we fix the width of the junction to $w=2a$ and the length to $W=42a$. This length ensures that the separation between edges and vortices is much larger than their respective localization lengths. The effective gap $\Delta_\mathrm{J}$ inside the junctions is estimated numerically from the spectrum of an infinitely long junction (see Fig.~\ref{fig:setup}), which yields $\Delta_\mathrm{J}\approx 0.12\Delta_0$.

 In the TSC, $\vartheta(\boldsymbol{x},t)=\eta(\boldsymbol{x})+\varphi(\boldsymbol{x},t)$ is the pair potential phase
with $\eta$ describing the vortices by $\nabla\times\nabla\eta=\sum_{\boldsymbol{x}_\mathrm{vortex}} 2\pi\delta(\boldsymbol{x}-\boldsymbol{x}_\mathrm{vortex})$; $\nabla\cdot\nabla\eta=0$, and $\varphi(\boldsymbol{x},t)$ describing the time-dependent bias, which is only nonzero in the middle superconductor and given by:
\begin{align}
    \varphi(t) = 2\pi\left( \theta(\tau-t) t/\tau + \theta(t-\tau) \right), \hspace{.5cm} t\geq 0
\end{align}
over a characteristic time $\tau$. Here $\theta(t)$ denotes the Heaviside step function. For this profile, the estimated injection time is simply $t_\mathrm{inj}=\tau \hbar v/(\Delta_\mathrm{J}W)$.  \\

\subsection{Computation of observables in the evolved many-body state}

Before the injection, the system is assumed to be in the stationary ground state of $\hat{H}(0)$ denoted by $\ket{\Omega}$. Here, we consider the evaluation of single-particle operators in the evolved many-body state $\hat{U}(t)\ket{\Omega}$ with the time-evolution operator $\hat{U}(t) = T\exp(-(i/\hbar)\int_0^t\hat{H}(t')\mathrm{d}t')$, $T$ being the time-ordering operator. Relative to the initial ground state, the net change in the expectation value of a single-particle operator $\hat{A}$ is denoted:
\begin{equation}
    \langle \hat{A}(t)\rangle - \langle \hat{A}(0)\rangle:=\bra{\Omega}\hat{U}^\dagger(t)\hat{A}\hat{U}(t)\ket{\Omega} - \bra{\Omega}\hat{A}\ket{\Omega} . 
    \label{eq:net_change}
\end{equation}

The effective description of the superconductor can be reduced to a non-interacting model using the BdG formalism. In App.~\ref{app:second_to_first_q}, we show how we can transform this many-body problem into single-particle problems which can be solved within the first quantization formalism.
Eq.~\eqref{eq:net_change} can be written as:
\begin{equation}
    \langle \hat{A}(t)\rangle - \langle \hat{A}(0)\rangle
    = \frac{1}{2}\sum_{ \alpha\in S^-}\Big(\bra{\alpha(t)}A\ket{\alpha(t)} - \bra{\alpha}A\ket{\alpha}\Big).
    \label{eq:obs_evolution}
\end{equation}
Here $A$ is the single-particle BdG operator associated with $\hat{A}$, $\ket{\alpha}:=\ket{\alpha (0)}$ denotes the $\alpha$-th eigenstate of $H(0)$ and $\ket{\alpha (t)}$ obeys
\begin{equation}
i\hbar\partial_t\ket{\alpha(t)} = H(t)\ket{\alpha(t)}.
\end{equation}
The evolution of the state $\ket{\alpha(t)}$ is calculated numerically using the python package Tkwant \cite{Gro13,Klo14,Wes14,Bau14,Ros18}. 
This approach has numerical complications as it requires to evolve all the $N$ states in $S^-$ in order to achieve convergence (see App.~\ref{app:convergence}).

We resolve this issue by writing $A$ in terms of the basis of eigenstates of $H(0)$:
\begin{equation}
    \langle \hat{A}(t)\rangle - \langle \hat{A}(0)\rangle =\Re \sum_{\substack{ \alpha\in S^-\\ \mu\in S^+}}\sum_{\nu\in S}\bra{\alpha (t)}\ket{\mu}\bra{\mu}A\ket{\nu}\bra{\nu }\ket{\alpha(t)} \label{eq:singleparticle_form}.
\end{equation}
Here the sets $S^+$ and $S^-$ denote positive and negative energy state indices respectively \footnote{Notice that particle-hole symmetry enforces that the eigenstates of the BdG Hamiltonian $H$ come in pairs of opposite energies. The eigenspace of zero modes of $H$ must be even dimensional and there must exist a basis of particle-hole partners in it. For each pair, we arbitrarily chose one state to be in $S^+$ and put its partner in $S^-$. Thus, in general, $S^+$ contains zero modes. Overall it contains half of the states ($n$ states) and if we act on them with the particle-hole symmetry operator, we obtain $S^-$.} and $S$ their union $S^+\cup S^-$.
In contrast with Eq.~\eqref{eq:obs_evolution} (see App.~\ref{app:convergence}), this form only gives non-zero contributions in a finite range around $E=0$. This allows us to approximate this expression by truncating the sum and discarding all terms above some energy cut-off, i.e terms with $| E_{\alpha,\mu,\nu}|>E_\mathrm{max}$.

\section{Results}
\label{sec:charge}

In this section we present the main results of our simulation. We show the charge signature of the braiding protocol and calculate the corresponding parity switch. We consider a system where $W$ is smaller, but comparable to the injection time $vt_\mathrm{inj} \approx 2W$. While the theoretical description, relying on the adiabatic limit, no longer holds for this system we show that the main predictions remain unchanged.

\subsection{Quantized charge measurement \label{subsec:quantized_charge}}

We first consider the charge signature that can be measured at the exit of the device, after the fusion of the edge vortices. For this we evaluate the current density operator $\hat{\jmath}_y(\boldsymbol{x})=(ev/a^2)\hat{\Psi}^\dagger(\boldsymbol{x})\nu_0\sigma_y\hat{\Psi}(\boldsymbol{x})$ in the $y$-direction using Eq.~\eqref{eq:singleparticle_form}. Here, $\nu_0$ is the identity acting on the particle-hole degree of freedom.
Defining the current as:
\begin{equation}
I(t)=a\sum_{\boldsymbol{x}|y=y_\mathrm{exit}}\langle\hat{\jmath}_y(\boldsymbol{x},t)\rangle-\langle\hat{\jmath}_y(\boldsymbol{x},0)\rangle
\label{eq:current_density_at_exit}
\end{equation}
the net charge creation is given by the time integral:
\begin{equation}
    Q(t)=\int_0^tI(t')\mathrm{d}t'.
\end{equation}
With this, we can calculate the charge pumped during the braiding protocol at the exit of the device ($y_\mathrm{exit}$).
The spatial separation $L$ between the two Josephson junctions allows to distinguish between two characteristic charge signatures. When $L\gg vt_\mathrm{inj}$, the injection events at each junction are well separated in space. In this case, the two pairs of edge-vortices produce separate signals of $\pm\frac{e}{2}$ charge at the exit. The charge contribution of the second pair of edge vortices experiences a sign flip in the presence of bulk vortices, as a consequence of braiding \cite{Ada09}. The theoretical predictions from Refs.~\onlinecite{Bee08,Ada09} are compared with numerical results in the left panel of Fig.~\ref{fig:charge}. On the other hand when $L\lesssim vt_\mathrm{inj}$, the injection events at both junctions are close, so that the overlapping electrical signals add up, producing a unit charge signature (Fig.~\ref{fig:charge} (b)). 

The transferred charge is an indirect probe of the braiding event as it is a result of the fusion between the edge vortices. It is therefore only quantized if the path lengths of the two vortices between injection and fusion are the same \cite{Ada09}. In contrast, the parity exchange is topologically protected, it does not depend on microscopic details. We will check this numerically.

\begin{figure}
    \includegraphics[width=\linewidth]{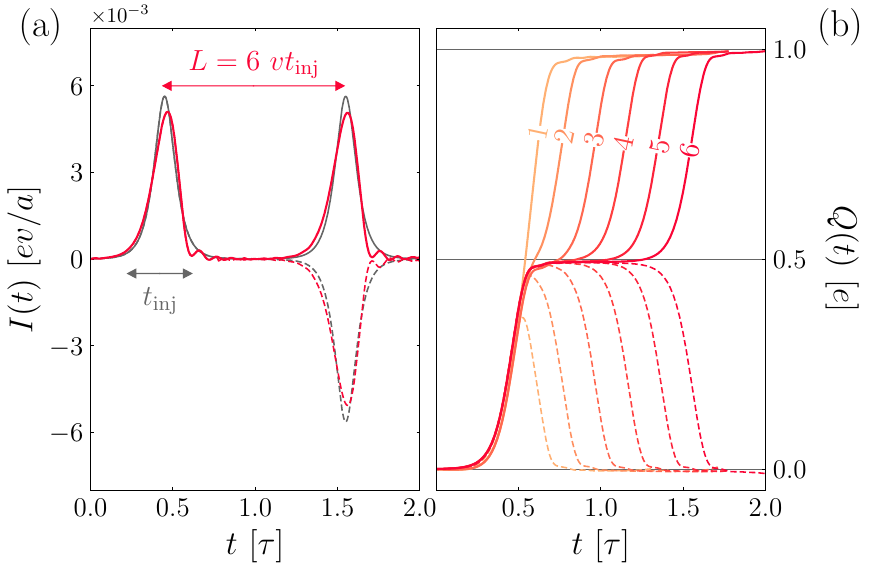}
    \caption{(a) Simulated (pink) and theoretical (gray) current density at the exit of the superconductor. A system without (with) vortices is represented with dashed (solid) lines. The pulse width $t_\mathrm{inj}\approx\tau/5.17$ is indicated. (b) Corresponding charge increase, for different values of the inter-junction separation $L$, with values of $L/v\tau$ shown on top of the curves. All simulations have $\tau=500a/v$ and $W=42a$. }
    \label{fig:charge}
\end{figure}

\subsection{Parity switch of edge-vortices}
\label{sec:parity}

The phase rotation $\varphi(t):0\rightarrow2\pi$ in the superconductor changes the parity locally carried by the two bulk vortices. Since parity must be globally conserved, then necessarily there must be an odd number of excitations elsewhere in the system -- namely carried by the edges. \cite{Bee08}. This change of parity  is a direct consequence of braiding between the bulk and edge vortices. To characterize this process we first identify the parity subsectors that correspond to the states in the bulk vortices and the edges. 

The full parity operator can be written --up to the sign of the initial ground state parity-- in terms of the Bogoliubov operators as:
\begin{equation}
    \hat{P} = \prod_{\alpha \in S^+}\left(1-2d^\dagger_\alpha d_\alpha\right)
    \label{eq:bdg_parity_operator}.
\end{equation}
We provide a further explanation for this form in App.~\ref{app:parity}. In our device, $\hat{P}$ can be split in a product of two terms, the first one corresponding to the bulk vortex excitation (i.e. the fermionic superposition of the two vortex Majorana zero-modes) and the second one containing all other excitations:
\begin{equation}
\begin{split}
    \hat{P} &= \left(1-2d^\dagger_{\alpha_v} d_{\alpha_v}\right) \cdot\prod_{\substack{\alpha \in S^+\\\alpha\neq\alpha_v}}\left(1-2d^\dagger_\alpha d_\alpha\right)\\
    &:=\hat{P}_{\text{vortices}}\cdot\hat{P}'
\end{split}\label{eq:bdg_parity_splitting}
\end{equation}
where $\alpha_v$ is the index of the fermionic state bound to the vortices.
This can be done if the vortex state is well isolated from the rest (i.e. there is no hybridization between vortex and edge states). $\hat{P}'$ can be evolved in the Heisenberg picture and expressed in terms of the Bogoliubov operators of the initial Hamiltonian $\{d_\beta\}_{\beta \in S}$. As we show in App.~\ref{app:parity}, the time evolution of each $d_\alpha$ can be expanded as
\begin{equation}
    \hat{U}^\dagger d_\alpha\hat{U} = \sum_{ \beta \in S}\chi_{\alpha\beta}d_\beta\quad\text{with} \quad\chi(t)_{\alpha\beta} =\bra{\alpha(0)}\ket{\beta(t)}
\end{equation}
The time evolution of $\hat{P}'$ can then be expressed as a sum of terms of different orders in $d$ operators
\begin{equation}
\begin{split}
    &\hat{U}^\dagger \hat{P}'\hat{U} =\Big(1-2\sum_{{ \alpha\in S^+}}\sum_{\mu,\nu \in S}\chi^*_{\alpha\mu}\chi_{\alpha\nu}d^\dagger_\mu d_\nu\\&+4\sum_{\substack{\alpha,\beta\in S^+\\E_\beta>E_\alpha}} \sum_{\mu,\nu,\sigma,\tau \in S}\chi^*_{\alpha\mu}\chi_{\alpha\nu}\chi^*_{\beta\sigma}\chi_{\beta\tau}d^\dagger_\mu d_\nu d^\dagger_\sigma d_\tau+\cdots\Big).
\end{split}
\end{equation}
Its expectation value in the ground state $\ket{\Omega}$ can then be calculated making use of Wick's theorem up to all orders. The final equation can be found in App.~\ref{app:parity} (Eq.~\eqref{eq:parity_evolution}).

In our numerical calculation we neglect correlators of order higher than four, and only include states within an energy window $E_\mathrm{max}$. This energy window is chosen to match the maximum excitation energy in order for the parity calculation to converge (see App.~\ref{app:exhaustive}).

Since edge and junction states are hybridized, $\hat{P}'$ cannot be decomposed similarly in edge and junction sectors. However, after the bias pump, the expectation value $\langle\hat{P}'\rangle$ can be identified with the parity carried by the edges $\langle\hat{P}_\mathrm{edge}\rangle$ as long as the filling of junction states -- which only exist for energies $E \geq \Delta_\mathrm{J}$ -- is negligible. The different intensities of red in Fig.~\ref{fig:parity} show the value obtained for $\hat{P}'$ as we increase $E_\mathrm{max}$. We see that convergence is achieved before we need to include any states with energies around $\Delta_J$. This identification of $\langle\hat{P}'\rangle\approx\langle\hat{P}_\mathrm{edge}\rangle$ is further supported in Sec.~\ref{sec:bound_states} and App.~\ref{app:exhaustive}.

Fig.~\ref{fig:parity}, shows that the parity expectation of the edges is unchanged when there are no vortices, but it switches in the presence of bulk vortices. This demonstrates that, for this set of parametes, the braiding of edge-vortices holds dynamically, and that the internal degrees of freedom in the junction do not spoil the exchange of parity. This implies that neither the adiabatic nor the point junction limits need to be satisfied for braiding to be realised.

\begin{figure}
    \includegraphics[width=\linewidth]{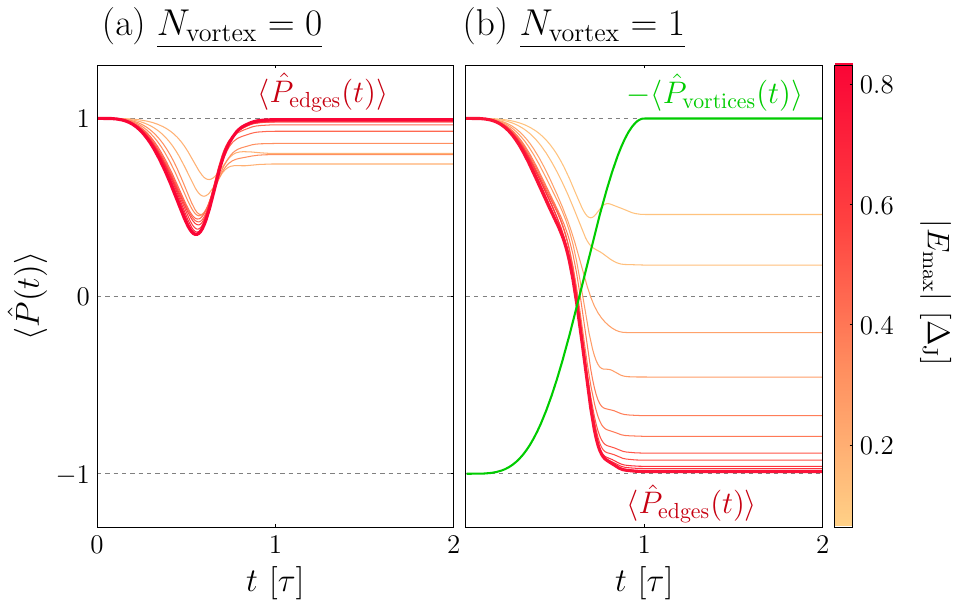}
	\caption{Evolution of the parity operator expectation value in the initial ground state without (a) and with vortices (b). In panel (b), the parity of the vortices is separated from the edges, and a parity switch is observed. Convergence of the curves as a function of $E_\mathrm{max}$ is shown in color.}
	\label{fig:parity}
\end{figure}

\subsection{Topological protection of the edge vortices}

The phase domain wall created during the quench corresponds to a pair of edge vortices that propagate along the edges. As one of them surrounds the bulk vortex it picks up a phase that realises the parity switch \cite{Ada09}. Since a $\pi$ domain wall cannot be unwound, this mechanism is protected from all local sources of disorder. In this part, we verify that the dynamically injected vortices are topologically protected by introducing irregularities in the spatial profile of $\Delta_0(\boldsymbol{x})$. We show how an additional path-length $\delta x$ in the upper edge (see the top panel of Fig.~\ref{fig:parity_charge}) influences the charge signature, fully spoiling the quantization discussed in Sec.~\ref{subsec:quantized_charge} in agreement with the predictions in Ref.~\onlinecite{Ada09}. In contrast, our calculation of parity (see the bottom panel of Fig.~\ref{fig:parity_charge}) remains unaffected by the local changes in the system, demonstrating the topological protection of the edge-vortex excitations. This confirms that even for a finite junction, edge-vortices can be used to encode protected quantum information.

\begin{figure}
        \includegraphics[width=\linewidth]{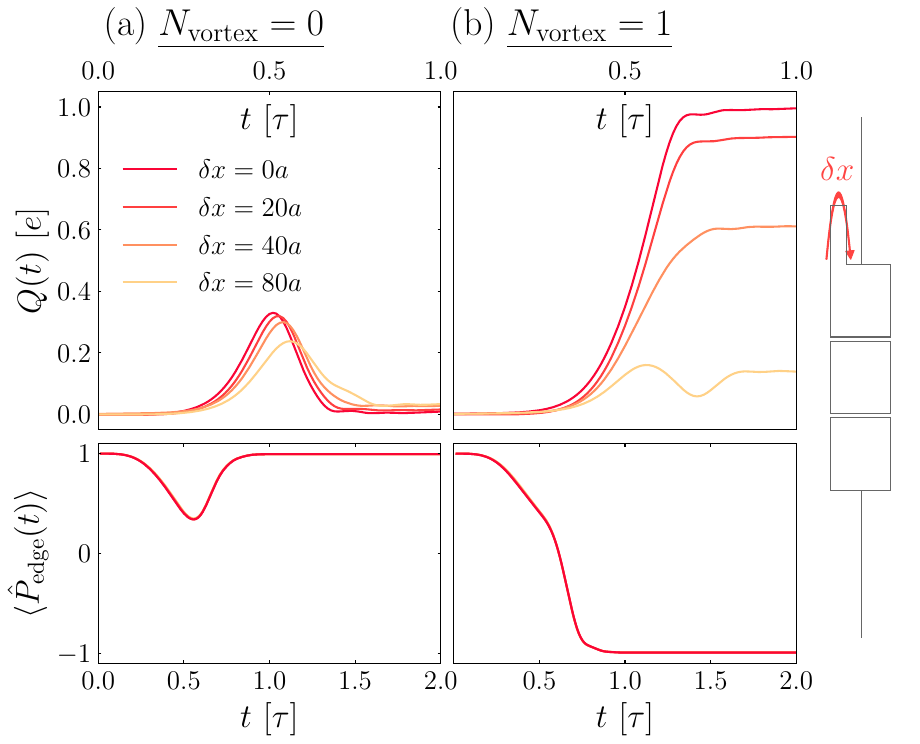}
	\caption{(Top) Net charge increase at the exit of the superconductors without (a) and with (b) vortices, with four geometrically induced path length differences between the edges $\delta x$ for $\tau=500a/v$. (Bottom) Parity of the edge sector for the same data sets. The calculated parity is independent of $\delta x$.  In this case the data sets overlap making the different curves indistinguishable.}
	\label{fig:parity_charge}
\end{figure}

\section{Long junction dynamics}
\label{sec:bound_states}
Our results so far have considered the particular case $vt_\mathrm{inj}\sim2W$ where the injection process is not spoiled by the excitation of junction modes. In this section, we consider the more general case where the ratio $vt_\mathrm{inj}/W$ is varied. In particular, we investigate how trapped excitations can influence the creation of edge-vortices for sufficiently long-junctions.

\subsection{Quasi-particle excitation spectrum}
To understand the behaviour in the junction we first study the quasi-particle excitation spectrum $E(\varphi)$. Within the superconducting gap, this spectrum consists of states localized in the bulk vortices, junction and edges. The injection process is characterized by the gap closing at $\varphi=\pi$ with the dispersion $E_\mathrm{J}=\pm\Delta_\mathrm{J}\cos\varphi/2$ seen before in Fig.~\ref{fig:setup}. In our case, the junction states couple with the edge states, forming hybridized bands seen in Fig.~\ref{fig:excitation_spectrum} (gray lines). We calculate the occupation number of these energy levels:
\begin{equation}
    \hat{N}(\varphi)=\sum_{ E_\mu(\varphi)\in S^+} d^{\dagger}_{\mu}(\varphi)d_{\mu}(\varphi) \label{eq:bdg_particle_number}
\end{equation}
where each term $d^{\dagger}_{\mu}(\varphi)d_{\mu}(\varphi)$ counts the quasi-particle occupation within a single energy level $\mu$. The expectation value in the evolved state $\hat{U}(t)\ket{\Omega}$ is then given by:
\begin{equation}
    \langle \hat{N}(\varphi,t)\rangle = \Re\sum_{\substack{ \alpha\in S^-\\   \mu\in S^+}}\sum_{\nu \in S} \bra{\alpha (t)}\ket{\mu^{\varphi}}\bra{\mu^{\varphi}}N\ket{\nu^{\varphi}}\bra{\nu^{\varphi}}\ket{\alpha(t)} \label{eq:n_phi}
\end{equation}
where $\ket{\mu^{\varphi}}$ denotes an eigenstate of $H(\varphi)$ and $N = \mathds{1}$. 

The occupation of each level through-out the quench is shown by thick lines in Fig.~\ref{fig:excitation_spectrum}, where the color is used to distinguish between edge (red) and junction (blue) states \footnote{The color at a value $\varphi$ and band $\mu$ is proportional to the value $\sum_{\boldsymbol{x}\in\mathrm{junctions}}|\langle\mu^\varphi|\boldsymbol{x}\rangle|^2$}. The slow injection case (a) treated in Sec.~\ref{sec:charge} shows that the junction states are only occupied near values of $\varphi=\pi$ and fully emptied in the edges at the end of the injection. In panel (b), the injection is short enough to create excitations in the levels $E>\Delta_\mathrm{J}$. Note that, in this case, the approximation $\langle\hat{P}'\rangle$ made in Sec.~\ref{sec:parity} fails because of nonzero occupation in the junction. This means that the parity switch is no longer fully carried by the edge modes, which we attribute to trapped excitations in the Josephson junction.

\begin{figure}
    \centering
    \includegraphics[width=\linewidth]{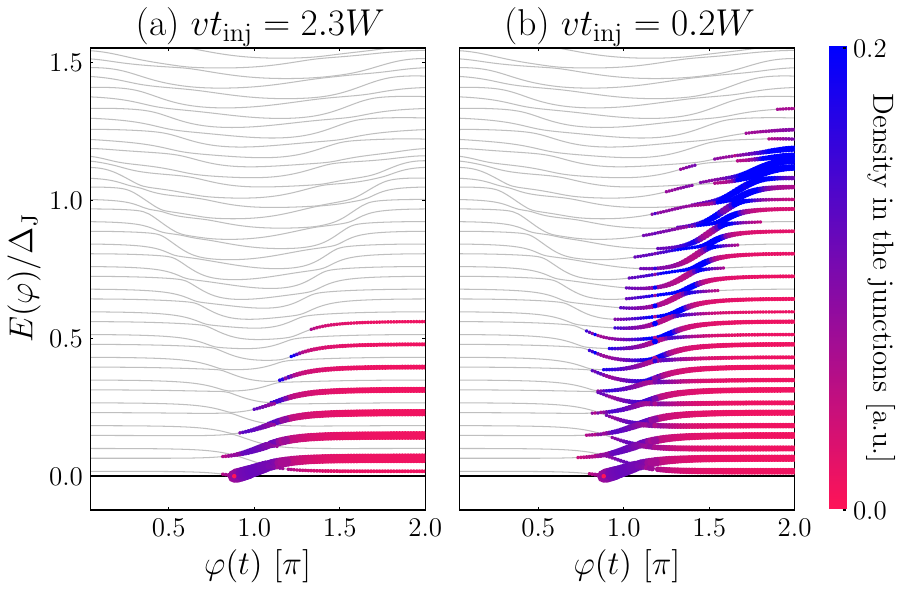}
    \caption{\color{black} Quasi-particle occupation of the energy levels (thick colored lines; a thick line signifies a strong occupation) above the ground state level ($E=0$), superimposed on the time-independent energy spectrum of $H(\varphi)$ (thin gray lines). The color of the lines distinguishes between junction (blue) and edge (red) states. At fast injection (b), the quasiparticle occupation in the junction levels $E_\mu\geq\Delta_\mathrm{J}$ at final time is high. We have removed the vortex state from this figure.}
    \label{fig:excitation_spectrum}
\end{figure}

\subsection{Trapped excitations}
In the presence of a finite Josephson junction the coupling between the two edges is mediated by their hybridization with the chiral states in the Josephson junction. This  hybridization is only supported for a duration $t_\mathrm{inj}$ around $\varphi=\pi$, when the junction is effectively gapless. We have shown that 
when $vt_\mathrm{inj}\sim 2W$ the travel time $W/v$ is short enough to allow the excitations to escape the junctions before the gap re-opens. Here we show that in the alternative regime $vt_\mathrm{inj}<W$, the excitation is partially trapped in the gapped bound state of the junction.

In order to describe the quasi-particles inside the junction, we define an excitation density via a spatial projection of the quasi-particle number $N(\boldsymbol{x})=\mathcal{P}(\boldsymbol{x}) N \mathcal{P}(\boldsymbol{x})$.
This is done similarly to our description of charge (i.e. $\bra{\boldsymbol{x}'}N(\boldsymbol{x})\ket{\boldsymbol{x}''}= \sigma_0\nu_0\delta_{\boldsymbol{x}',\boldsymbol{x}''}\delta_{\boldsymbol{x},\boldsymbol{x}'}$) arriving to the expression:
 \begin{equation}
    \langle\hat{\rho}_\varphi(x,t)\rangle = \Re\sum_{\substack{ \mu\in S^+\\  \alpha\in S^-}} \sum_{\nu \in S}\bra{\alpha (t)}\ket{\mu^{\varphi}}\bra{\mu^{\varphi}}N(\boldsymbol{x})\ket{\nu^{\varphi}}\bra{\nu^{\varphi}}\ket{\alpha(t)} \label{eq:local_excitation_density}.
\end{equation}
 Note that when integrated over the whole system, the Eq.~\eqref{eq:n_phi} is recovered. Integrating this density locally gives the number of quasi-particle inside junctions $\langle\hat{N}_\mathrm{junc}(t)\rangle$ and edges $\langle\hat{N}_\mathrm{edges}(t)\rangle$. 

\begin{figure}
    \centering
    \includegraphics[width=\linewidth]{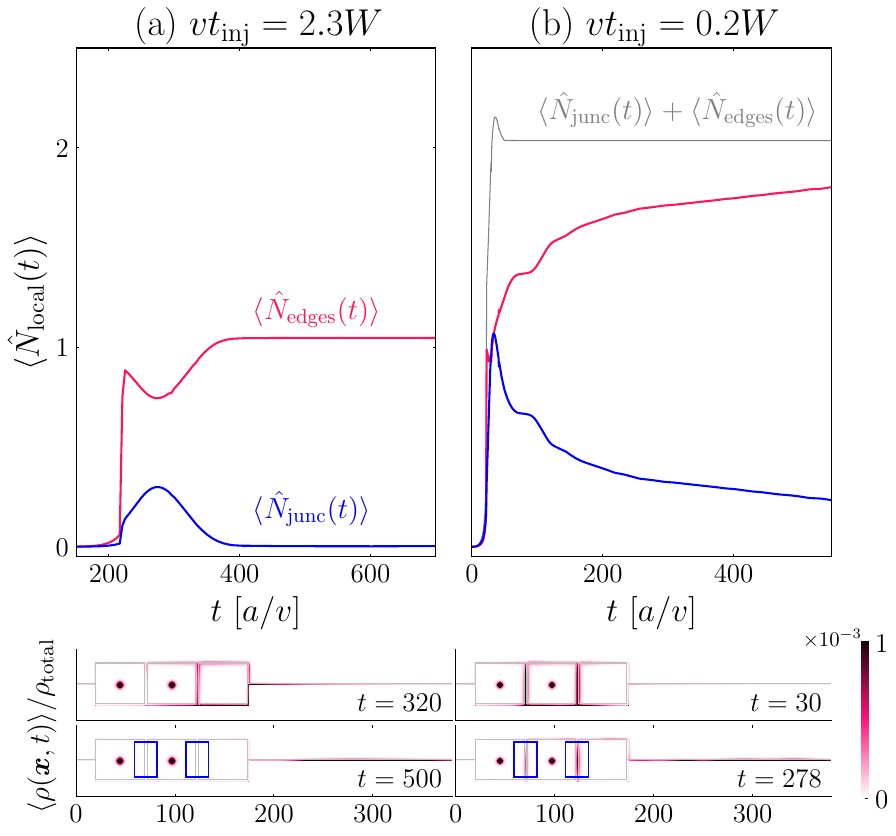}
    \caption{\color{black}Bogoliubov quasi-particle number inside the junction (red) and inside the edges (blue) as a function of time. Panel (a) shows that the junction excitation fully escapes into the edges, while in (b), at short injection time, the junction contains residual quasi-particles. The bottom panels show the corresponding quasi-particle densities at two different times times. The integration window used to calculate the quasi-particle number inside the junction is marked with a blue rectangle. An animated visualisation can be found in \cite{supplemental}.}
    \label{fig:excitation_density}
\end{figure}

In Fig.~\ref{fig:excitation_density}, we show how the quasi-particle changes with time for two different systems. When the injection is slow (a) the quasi-particle number in the junction is fully transferred to the edges as anticipated. In the alternate case when the injection is very fast (b), the particle number slowly decays towards a constant residual value in the junctions corresponding to quasi-particles occupying the lowest bound state in the Josephson junctions. As this trapped excitation can carry a part of the parity exchange it can spoil the injection protocol as well as the characteristic charge signature (shown in App.~\ref{app:exhaustive}). For this reason it is important to find a bound when the trapped excitations in the junction can be neglected.

\subsection{Particle number in the junction}
In the adiabatic theory Ref.~\onlinecite{Bee08}, the total particle number produced in the edges at final time is equal to $1.037$. The non-quantized number is due to particle-hole pairs production during the injection process. At slow injection, we find a comparable value $\langle\hat{N}_\mathrm{junc}\rangle+\langle\hat{N}_\mathrm{edge}\rangle=1.049$ as indicated in Fig.~\ref{fig:excitation_density} (a), close to the adiabatic theory. For the fast injection in Fig.~\ref{fig:excitation_density}, this is $\langle\hat{N}_\mathrm{junc}\rangle+\langle\hat{N}_\mathrm{edge}\rangle=2.033$ instead.

We therefore turn to a quantitative description of the residual particle number in the junction $\langle\hat{N}_\mathrm{junc}\rangle$ for different values of $vt_\mathrm{inj}/W$. We achieve this by simulating different values of $\tau$ in Fig.~\ref{fig:junction_density}. In Fig.~\ref{fig:junction_density} (a), the particle number is shown as a function of time for different values of $vt_\mathrm{inj}/W$, where we distinguish between the two regimes $vt_\mathrm{inj}>W$ and $vt_\mathrm{inj}<W$ by two colors. In panel (b), we show that the residual excitation number in the junction decreases fast as the injection time becomes long. We match this with an exponential shown in Fig.~\ref{fig:junction_density}. After $vt_\mathrm{inj}>2W$, this value has nearly decayed to zero. In an experimental setting, this provides us with an upper bound on the flux bias change rate $|\mathrm{d}\Phi/\mathrm{d}t|<\Phi_0v/2W^2\Delta_\mathrm{J}$ when the parity exchange is fully carried by the edges corresponding, ensuring a successful injection of edge vortices. \\

\begin{figure}
    \centering
    \includegraphics[width=\linewidth]{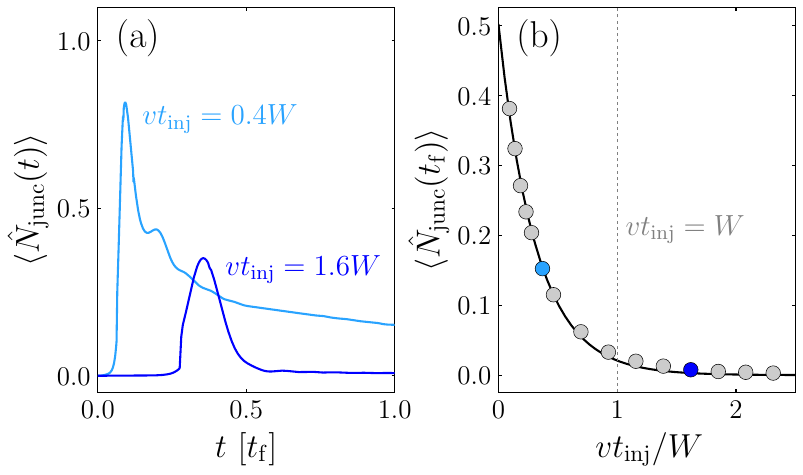}
    \caption{\color{black}(a) Quasi-particle number inside the junction as a function of time, for two values of the injection time. (b) Residual quasi-particle number in the junction at some final time $t_f=500a/v$ as a function of the ratio $vt_\mathrm{inj}/W$. An exponential fit yields $\langle\hat{N}_\mathrm{junc}(t_\mathrm{f})\rangle=N_0\cdot\exp(vt_\mathrm{inj}/W\beta)$ with $\beta=0.31$ and $N_0=0.5$.}
    \label{fig:junction_density}
\end{figure}

\section{Conclusion}
\label{sec:conclusion}
In this work we have shown how a braiding protocol introduced in Ref.~\onlinecite{Bee08} can be dynamically simulated as a tight-binding many-body system. With this setup we were able to fully probe the braiding process away from the limitations of the effective model. This allowed us to investigate the relevant scales in the system as well as compare the current signature with analytical predictions. We were able to study dynamically the local parity switch present in the edge states and show the topological protection of this exchange. We have shown that the injection and braiding of edge-vortices is uncompromised by a finite junction when $vt_\mathrm{inj}>2 W$, so that all the parity exchange is contained in the edge states. Additionally we studied this system away from this limit and investigated the excitations in the junction. Here, we showed that the lowest bound state of the junction remains excited long after the quench for sufficiently fast injections. While the parity switch $\langle \hat{P}'\rangle$ is still protected in this limit, we can no longer conclude that it is fully carried in the edge states, therefore providing a limitation for the use of such device as a topological qubit. For this reason we show the interplay of scales $vt_\mathrm{inj}, W$ to find a parameter regime, where the injection of edge vortices is well defined. We see that the adiabatic condition $vt_\mathrm{inj}\gg  W$ discussed in previous works can be relaxed into $vt_\mathrm{inj}\gtrsim  W$, while keeping the braiding predictions intact. This is helpful for future experimental work as it allows large deviations from the point junction limit.\\

\acknowledgments	
We thank A. R. Akhmerov and \.{I}. Adagideli for helpful discussions. This project has received funding from the European Research Council (ERC) under the European Union's Horizon 2020 research and innovation program.


\appendix
\section{Time-evolution of single-body operators in BdG}

\subsection{From second to first quantization}
\label{app:second_to_first_q}
In a tight-binding system, any single-body operator $\hat{A}$ can be written as
\begin{equation}
    \hat{A} = \sum_{\alpha,\beta =1}^{n}A^e_{\alpha\beta}\hat{\psi}^\dagger_\alpha \hat{\psi}_\beta,\qquad A^e_{\alpha\beta} = \bra{0}\hat{\psi}_\alpha\hat{A}\hat{\psi}^\dagger_\beta\ket{0},
\end{equation}
where $\ket{0}$ denotes the vacuum of electrons, which can be rewritten into the BdG form as
\begin{equation}
         \hat{A} = \frac{1}{2}{\hat{\Psi}}^\dagger A \hat{\Psi}+\frac{1}{2}\Tr A^e\qquad
\end{equation}
with
\begin{equation}
\begin{split}
        &A=\begin{pmatrix}
                    A^e&0\\
                    0&-\sigma_yA^{e*}\sigma_y
                    \end{pmatrix}\\
        & \begin{matrix}\hat{\Psi} := \Big(\hat{\psi}_{1\uparrow} & \hat{\psi}_{1\downarrow}\cdots & \hat{\psi}_{N/2\uparrow} & \hat{\psi}_{N/2\downarrow}\\ & \hat{\psi}^\dagger_{1\downarrow} & -\hat{\psi}^\dagger_{1\uparrow}\cdots & \hat{\psi}^\dagger_{N/2\downarrow} & -\hat{\psi}^\dagger_{N/2\uparrow} \Big)^T\end{matrix}
\end{split}
    \label{eq:first_q_operator}
\end{equation}
We can evolve this operator in the Heisenberg picture to obtain
\begin{equation}
    \hat{U}^\dagger\hat{A}\hat{U} = \frac{1}{2}\hat{\Psi}(t)^\dagger A \hat{\Psi}(t)+\frac{1}{2}\Tr A^e 
\end{equation}
where we defined $\hat{\psi}_\alpha(t) = \hat{U}^\dagger \hat{\psi}_\alpha\hat{U}$. Since we intend to evaluate this operator in the ground state $\ket{\Omega}$ of the initial Hamiltonian, we need to write it in terms of the Bogoliubov operators $\{d_\beta\}_{\nu \in S}$ of $\hat{H}(0)$. It is possible to prove (see App.~\ref{app:proof_time_evolution}) that the $\{\hat{\psi}_\alpha(t)\}_{\nu \in S}$ operators can be written as linear combinations of these Bogoliubov operators as
\begin{equation}
    \hat{\psi}_\alpha(t) = \sum_{\beta\in S}\Phi_{\alpha\beta}(t)d_\beta \qquad \text{i.e.} \qquad \hat{\Psi}(t) = \Phi(t) \mathbf{d}
\end{equation}
where $\Phi(0)$ is the matrix that diagonalises the BdG Hamiltonian at $t = 0$, (i.e. $H(0) =\Phi(0)\mathcal{E}\Phi^\dagger(0)$) and $\Phi(t)$ is the solution of
\begin{equation}
   i\hbar\partial_t\Phi(t)  = H(t)\Phi(t).
\end{equation}
Notice that this means that the columns of $\Phi$ are none other than the eigenstates of $H(0)$ evolved according to the Schr\"odinger equation for $H(t)$. 
With this, we can express
\begin{equation} 
    \hat{U}^\dagger\hat{A}\hat{U} = \frac{1}{2}\mathbf{d}^\dagger \Phi^\dagger A \Phi\mathbf{d}+\frac{1}{2}\Tr A^e
\end{equation}
Finally, using the fact that by definition $ \bra{\Omega}d_\alpha^\dagger d_\beta\ket{\Omega} = \delta_{\alpha\beta}$ if $ E_{\alpha} < 0$ and $ \bra{\Omega}d_\alpha^\dagger d_\beta\ket{\Omega} = 0$ otherwise, we obtain
\begin{equation}
    \langle \hat{A}(t)\rangle - \langle \hat{A}(0)\rangle = \frac{1}{2}\sum_{\alpha\in S^-}\left(\Phi^\dagger(t) A\Phi(t) - \Phi^\dagger(0) A\Phi(0)\right)_{\alpha\alpha},
    \label{eq:first_q_equation}
\end{equation}
which in Dirac notation becomes 
\begin{equation}
    \langle \hat{A}(t)\rangle - \langle \hat{A}(0)\rangle
    = \frac{1}{2}\sum_{ \alpha\in S^-}\Big(\bra{\alpha(t)}A\ket{\alpha(t)} - \bra{\alpha(0)}A\ket{\alpha(0)}\Big).
    \label{eq:obs_evolution_appendix}
\end{equation}
With this, we have mapped our original problem of evolving many-body states in a Hilbert space of dimension $2^n$ into $n$ first quantization problems in a Hilbert space of dimension $2n$.

\subsection{Convergence}
\label{app:convergence}
The fact that Eq.~\eqref{eq:obs_evolution_appendix} involves all $n$ negative energy eigenstates of $H$ poses two problems. First, we only aim at describing  the system accurately at low energies. Any realistic system will not share the specific high-energy behaviour of our tight-binding description far from the Fermi energy. Secondly, we should be able to understand our system by considering only states close to the Fermi energy, so evolving all of them is a waste of computational resources. Unfortunately we have no reason to belive that the contribution of both terms in Eq.~\eqref{eq:obs_evolution_appendix} will cancel out as we go away from the Fermi energy. This was actually studied numerically and it was verified that the value of $\langle \hat{\jmath}_y(\boldsymbol{x},t)\rangle-\langle \hat{\jmath}_y(\boldsymbol{x},0)\rangle$ as given by Eq.~\eqref{eq:obs_evolution_appendix} does not converge --instead it oscillates-- as we increase the amount of states evolved (see Fig.~\ref{fig:convergence_charge}). This section is devoted to rewrite this equation in a form that solves this issue. To do so, let us explicitly make use of basis of the eigenstates of $H(0)$ and introduce the completeness relation around $A$ in the first term of Eq.~\eqref{eq:obs_evolution_appendix} to obtain
\begin{equation}
\begin{split}
    \frac{1}{2} \sum_{ \alpha\in S^-}&\bra{\alpha(t)}A\ket{\alpha(t)} = \\
    \frac{1}{2} \sum_{ \alpha\in S^-}\sum_{\mu,\nu \in S}&\bra{\alpha(t)}\ket{\mu}\bra{\mu}A\ket{\nu}\bra{\nu}\ket{\alpha(t)} =\\
    \frac{1}{2} \sum_{ \alpha\in S^-}\sum_{ \mu,\nu\in S^-}\quad
    \Big(& \bra{\alpha(t)}\ket{\mu}\bra{\mu}A\ket{\nu}\bra{\nu}\ket{\alpha(t)} \\[-2ex]
    +&\bra{\alpha(t)}\ket{\mathcal{C}\mu}\bra{\mathcal{C}\mu}A\ket{\nu}\bra{\nu}\ket{\alpha(t)}\\[1ex]
    +&\bra{\alpha(t)}\ket{\mu}\bra{\mu}A\ket{\mathcal{C}\nu}\bra{\mathcal{C}\nu}\ket{\alpha(t)}\\
    +&\bra{\alpha(t)}\ket{\mathcal{C}\mu}\bra{\mathcal{C}\mu}A\ket{\mathcal{C}\nu}\bra{\mathcal{C}\nu}\ket{\alpha(t)}\Big).
\end{split}
\label{eq:expanded}
\end{equation}
where $\mathcal{C}=\sigma_y\nu_y\mathcal{K}$ is the BdG charge conjugation operator and $\mathcal{C}\mu$ denotes the particle-hole partner of the state labeled $\mu$. Since $\hat{A}$ is a single-particle operator, it satisfies $\mathcal{C}A\mathcal{C} = -A$. Given that $\{\ket{\alpha(t)}: \alpha\in\left(S^-\cup S^+\right)\}$ is a complete basis of the BdG Hilbert space, we can write the first term of Eq.~\eqref{eq:expanded} as
\begin{equation}
\begin{split}
    \frac{1}{2} \sum_{ \alpha\in S^-}\sum_{\mu,\nu\in S^-}&\bra{\alpha(t)}\ket{\mu}\bra{\mu}A\ket{\nu}\bra{\nu}\ket{\alpha(t)}=\\
    &\frac{1}{2} \sum_{\mu\in S^-}\bra{\mu}A\ket{\mu}\\ -\frac{1}{2} \sum_{ \alpha\in S^+}\sum_{ \mu,\nu\in S^-} &\bra{\mathcal{C}\alpha(t)}\ket{\mu}\bra{\mu}A\ket{\nu}\bra{\nu}\ket{\mathcal{C}\alpha(t)}
\end{split}
\label{eq:first_term}
\end{equation}
If we plug this in Eq.~\eqref{eq:expanded} and then in Eq.~\eqref{eq:obs_evolution_appendix}, a few simplifications happen. The first term of this equation will cancel with the second term of Eq.~\eqref{eq:obs_evolution_appendix}, and the second term of Eq.~\eqref{eq:first_term} is real and equal to the last term of Eq.~\eqref{eq:expanded} (this follows from the properties of $\mathcal{C}$). In addition, the second and third terms of Eq.~\eqref{eq:expanded} are each other's complex conjugate. Taking all of this into account we can write down Eq.~\eqref{eq:obs_evolution_appendix} as
\begin{equation}
    \begin{split}
        \langle \hat{A}(t)\rangle - \langle \hat{A}(0)\rangle = &\\
    \Re\sum_{ \alpha\in S^-}\sum_{ \mu,\nu\in S^+}\Big(&\bra{\alpha(t)}\ket{\mu} \bra{\mu}A\ket{\nu}\bra{\nu}\ket{\alpha(t)} \\[-2ex]
        +&\bra{\alpha(t)}\ket{\mu}\bra{\mu}A\ket{\mathcal{C}\nu}\bra{\mathcal{C}\nu}\ket{\alpha(t)}\Big)
    \end{split}
\end{equation}
which we write more simply in the main text as 
\begin{equation}
\langle \hat{A}(t)\rangle - \langle \hat{A}(0)\rangle =\Re \sum_{\substack{ \alpha\in S^-\\  \mu\in S^+}} \sum_{\nu \in S}\bra{\alpha (t)}\ket{\mu}\bra{\mu}A\ket{\nu}\bra{\nu }\ket{\alpha(t)}
\label{eq:obs_evolution_convergent}
\end{equation}
This formula includes overlaps between positive energy and evolved negative energy states which ensures non-zero contributions to only exist around $E=0$. In Fig.~\ref{fig:convergence_charge} we show how the contribution of the terms in the sum vanishes as we go further away from the Fermi energy, which lets us avoid having to evolve all negative energy states.

\begin{figure}
    \centering
    \includegraphics[width=\linewidth]{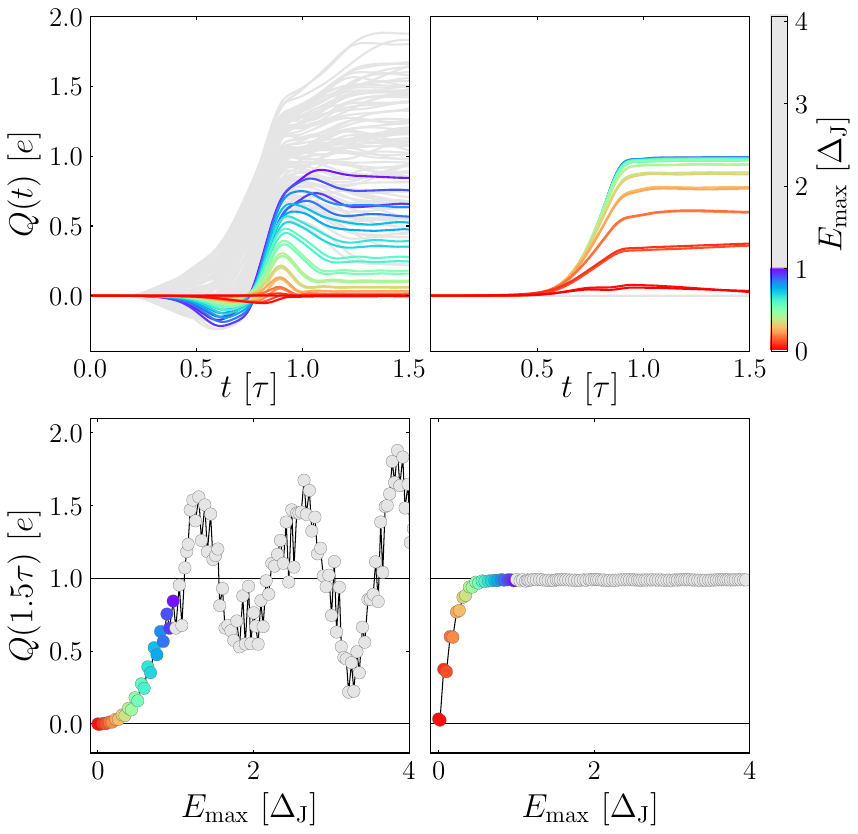}
    \caption{Convergence of the charge at the exit with two methods (Left) With charge expressed in the local basis using Eq.~\eqref{eq:first_q_equation} where $E_\mathrm{max}$ is the maximum energy of the states in the sum over $\alpha$.
    (Right) With charge expressed in the basis of eigenstates of $H(0)$ using Eq.~\eqref{eq:obs_evolution_convergent} where $E_\mathrm{max}$ is the maximum energy of the states in the sums over $\alpha$, $\mu$ and $\nu$.}
    \label{fig:convergence_charge}
\end{figure}

\subsection{Proof  of time evolution method}
\label{app:proof_time_evolution}
In this section we prove the following statement:
\begin{remark}
Let $\hat{\Psi}$ be the Nambu spinor of fermion creation and annihilation operators as defined in Eq.~\eqref{eq:first_q_operator} satisfying $\{\hat{\psi}_\alpha,\hat{\psi}^\dagger_\beta\} = \delta_{\alpha,\beta}$ and $\{\hat{\psi}_\alpha,\hat{\psi}_\beta\} = \delta_{\alpha,\mathcal{C}\beta}$ where $\mathcal{C}\alpha$ is the index of $(\hat{\psi}_\alpha)^\dagger$ in $\hat{\Psi}$ (i.e. $\hat{\psi}_{\mathcal{C}\alpha} = \hat{\psi}^\dagger_\alpha$). Let
$$\hat{H}(t) = \frac{1}{2}\hat{\Psi}^\dagger H(t)\hat{\Psi}$$ be the time-dependent BdG Hamiltonian describing a tight-binding superconducting system of fermions. Let $\hat{U}(t)$ be its corresponding evolution operator. Let $\mathcal{C}$ be the antiunitary charge conjugation operator satisfying $\mathcal{C}^2=1$ and $\{\mathcal{C},H\}=0$. Let $V(t)$ be a matrix that diagonalises $H(t)$ and let $\mathbf{d} = \left(d_1,d_2,\cdots,d_{2n}\right)$ be the spinor of Bogoliubov operators diagonalising $\hat{H}(0)$.

Then, the time evolution of $\hat{\Psi}$ can we written as
\begin{equation}
    \hat{\Psi}(t):=\hat{U}(t)^\dagger\hat{\Psi}\hat{U}(t) = \Phi(t) \mathbf{d}
    \label{eq:coef_definition}
\end{equation}
where $\Phi$ obeys
\begin{equation}
    i\hbar\partial_t\Phi(t)  = H(t)\Phi(t),\qquad\Phi(0) = V(0)
    \label{eq:to_prove}
\end{equation}
\end{remark}

\begin{proof}
 According to Heisenberg's picture evolution equation we have
\begin{equation}
    i\partial_t \hat{\psi}_\alpha(t) = \left[ \hat{\psi}_\alpha(t),\hat{U}(t)^\dagger \hat{H}(t) \hat{U}(t) \right].
    \label{eq:Heisenberg_evo}
\end{equation}
Since $\hat{H}$ is quadratic in $\hat{\Psi}$, we know that $\hat{\psi}_\alpha(t)$ can be expanded in terms of the initial $\hat{\psi}$'s as
\begin{equation}
    \hat{\psi}_\alpha(t) = \sum_{\beta}\zeta_{\alpha\beta}(t) \hat{\psi}_\beta,
    \label{eq:evo_expansion}
\end{equation}
or in matrix notation
\begin{equation}
    \hat{\Psi}(t)=\zeta(t)\hat{\Psi}.
\end{equation}
Notice that the unitarity of $\hat{U}$ imposes that the operators in $\hat{\Psi}(t)$ satisfy the same commutation algebra as the initial ones. In turn, this imposes unitarity on $\zeta$. We can use Eq.~\eqref{eq:evo_expansion} to write the commutator in Eq.~\eqref{eq:Heisenberg_evo} as
\begin{equation}
    \left[ \hat{\psi}_\kappa(t),\hat{U}(t)^\dagger \hat{H}(t) \hat{U}(t) \right] = 
   \frac{1}{2}\sum_{\alpha\beta\mu\nu\lambda}H_{\alpha\beta}
    \zeta^*_{\alpha\mu}\zeta_{\beta\nu}\zeta_{\kappa\lambda} [\hat{\psi}_\lambda,\hat{\psi}^\dagger_\mu \hat{\psi}_\nu]
\end{equation}
It is easy to check that
\begin{equation}
    [\hat{\psi}_\lambda,\hat{\psi}^\dagger_\mu \hat{\psi}_\nu] = \hat{\psi}_\nu\delta_{\lambda,\mu}-\hat{\psi}^\dagger_\mu\delta_{\lambda,\mathcal{C}\nu},
\end{equation}
so we get
\begin{equation}
\begin{split}
    \left[\hat{\psi}_\kappa(t),\hat{U}(t)^\dagger \hat{H}(t) \hat{U}(t) \right] &= 
     \frac{1}{2}\sum_{\alpha\beta\mu\nu}H_{\alpha\beta}\zeta^*_{\alpha\mu}\zeta_{\beta\nu}\zeta_{\kappa\mu} \hat{\psi}_\nu\\&-
      \frac{1}{2}\sum_{\alpha\beta\mu\nu}H_{\alpha\beta}\zeta^*_{\alpha\mu}\zeta_{\beta\nu}\zeta_{\kappa{\mathcal{C}\nu}} \hat{\psi}^\dagger_\mu. 
\end{split}
\end{equation}
Using $\hat{\psi}^\dagger_\mu = \hat{\psi}_{\mathcal{C}\mu}$ and relabeling in the last term we can rewrite
\begin{equation}
\begin{split}
    &\left[ \hat{\psi}_\kappa(t),\hat{U}(t)^\dagger \hat{H}(t) \hat{U}(t) \right] =\\&\frac{1}{2}\sum_{\alpha\beta\mu\nu}H_{\alpha\beta}\zeta^*_{\alpha\mu}\zeta_{\beta\nu}\zeta_{\kappa\mu} \hat{\psi}_\nu-
      \frac{1}{2}\sum_{\alpha\beta\mu\nu}H_{\alpha\beta}\zeta^*_{\alpha\mu}\zeta_{\beta\nu}\zeta_{\kappa,{\mathcal{C}\nu}} \hat{\psi}_{\mathcal{C}\mu}=\\
      &\frac{1}{2}\sum_{\alpha\beta\mu\nu}H_{\alpha\beta}\zeta^*_{\alpha\mu}\zeta_{\beta\nu}\zeta_{\kappa\mu} \hat{\psi}_\nu-
      \frac{1}{2}\sum_{\alpha\beta\mu\nu}H_{\alpha\beta}\zeta^*_{\alpha,{\mathcal{C}\nu}}\zeta_{\beta,{\mathcal{C}\mu}}\zeta_{\kappa\mu} \hat{\psi}_\nu.
\end{split}
\end{equation}
Comparing with the left-hand side of Eq.~\eqref{eq:Heisenberg_evo} we can deduce that
\begin{equation}
    i\partial_t\zeta_{\kappa\nu} =  \frac{1}{2}\sum_{\alpha\beta\mu}H_{\alpha\beta}\zeta^*_{\alpha\mu}\zeta_{\beta\nu}\zeta_{\kappa\mu}-
      \frac{1}{2}\sum_{\alpha\beta\mu}H_{\alpha\beta}\zeta^*_{\alpha,{\mathcal{C}\nu}}\zeta_{\beta,{\mathcal{C}\mu}}\zeta_{\kappa\mu}
\end{equation}
From $ \hat{\psi}^\dagger_\alpha(t)=\psi_{\mathcal{C}\alpha}(t)$, we have $\zeta_{\alpha\beta} = \zeta^*_{{\mathcal{C}\alpha},{\mathcal{C}\beta}}$ so the previous equation becomes
\begin{equation}
\begin{split}
    i\partial_t\zeta_{\kappa\nu} &=  \frac{1}{2}\sum_{\alpha\beta\mu}H_{\alpha\beta}\zeta^*_{\alpha\mu}\zeta_{\beta\nu}\zeta_{\kappa\mu}-
      \frac{1}{2}\sum_{\alpha\beta\mu}H_{\alpha\beta}\zeta_{{\mathcal{C}\alpha},{\nu}}\zeta^*_{\mathcal{C}\beta,{\mu}}\zeta_{\kappa\mu}\\
\end{split}
\end{equation}
The particle-hole symmetry of $H$ ($\mathcal{C}H\mathcal{C} = -H$) can be expressed element-wise as $H_{\mathcal{C}\alpha,\mathcal{C}\beta}=-H_{\beta,\alpha}$. After some relabeling on the last term, this lets us rewrite the previous equation as
\begin{equation}
    i\partial_t\zeta_{\kappa\nu} = \sum_{\alpha\beta\mu}H_{\alpha\beta}\zeta^*_{\alpha\mu}\zeta_{\beta\nu}\zeta_{\kappa\mu}
\end{equation}
The unitarity of $\zeta$ implies $\sum_\mu\zeta^*_{\alpha\mu} \zeta_{\kappa\mu}=\delta_{\alpha\kappa}$ so the previous expression becomes
\begin{equation}
   i\partial_t\zeta_{\alpha\beta} = \sum_{\mu}H_{\alpha\mu}\zeta_{\mu\beta}
   \label{eq:evo_lattice_operators}
\end{equation}
or in matrix notation
\begin{equation}
   i\partial_t\zeta  = H\zeta, \qquad\zeta(0) =\openone
   \label{eq:evo_lattice_operators}
\end{equation}
Now notice that we can compose Eq.~\eqref{eq:evo_expansion} with $\hat{\Psi} = V(0)\mathbf{d}$ and define $\Phi(t) = \zeta(t) V(0)$ that satisfies Eq.~\eqref{eq:coef_definition}. Since $V(0)$ is time-independent, Eq.~\eqref{eq:to_prove} follows immediately from  Eq.~\eqref{eq:evo_lattice_operators}.
\end{proof}

\section{Parity}
\subsection{Time evolution of the parity operator}
\label{app:parity}
The parity operator is defined as:
\begin{equation}
    \hat{P} = (-1)^{\sum_{\alpha = 1}^{n}{\hat\psi}^\dagger_\alpha \hat{\psi}_\alpha} = \prod_{\alpha=1}^{n}\left(1-2\hat{\psi}^\dagger_\alpha \hat{\psi}_\alpha\right).
\end{equation}
Since it commutes with $\hat{H}$, its ground state is an eigenstate of parity. This, together with the fact that the BdG operators switch the parity of a state, implies that we can also write down our parity operator in terms of them:
\begin{equation}
    \hat{P} = p_\Omega\prod_{\alpha \in S^+}^n\left(1-2d^\dagger_\alpha d_\alpha\right)
    \label{eq:bdg_parity_operator}
\end{equation}
where $p_\Omega=\pm1$ stands for the parity of the ground state. In general, we can express the parity of a set of quasi-particle states $S$ as
\begin{equation}
    \hat{P}_S = \prod_{\alpha \in S}\left(1-2d^\dagger_\alpha d_\alpha\right)
    \label{eq:parity_sector}
\end{equation}
The time evolution of this operator is given by substituting each $d_\alpha$ for $d_\alpha(t) = \hat{U}^\dagger d_\alpha\hat{U}$. From the results of App.~\ref{app:second_to_first_q}, it is straightforward to obtain the expression of $d_\alpha(t)$ in terms of $\{d_\alpha\}_{\alpha \in S}$:
\begin{equation}
    \hat{U}^\dagger\mathbf{d}\hat{U} = \hat{U}^\dagger V(0)^\dagger\hat{\Psi}\hat{U} = V(0)^\dagger\hat{\Psi}(t) = V(0)^\dagger\Phi(t) \mathbf{d}
\end{equation}
Thus, if we define $\chi(t) = V(0)^\dagger\hat{\Psi}(t)$ we have
\begin{equation}
    \hat{U}^\dagger d_\alpha\hat{U} = \sum_{\beta\in S}\chi(t)_{\alpha\beta}d_\beta\qquad\chi(t)_{\alpha\beta} =\bra{\alpha}\ket{\beta(t)}
\end{equation}

We can expand the product in Eq.~\eqref{eq:parity_sector} and use Wick's theorem to obtain an expression for the time evolution of $\langle\hat{P}_S\rangle$
\begin{equation}
         \langle\hat{P}_S(t)\rangle =\sum_{m = 0}^{n_S} (-2)^m\sum_{\substack{0< \alpha_1\\ <\dots< \alpha_m}}\sum_{c\in C_m}(-1)^{s(c)}\prod_{k = 1}^m\Theta^{X_k(c)Y_k(c)}_{\alpha_{i_k(c)}\alpha_{j_k(c)}}.
         \label{eq:parity_evolution}
\end{equation}
This formula contains several elements. First, we have a sum over all orders $0<m<n_S$ (the term corresponding to $m=0$ is equal to $1$). For each order $m$ we sum over all unordered choices of $m$ states among $n_S$.  For every such choice, we sum over all possible Wick contractions of that order ($C_m$ denotes the set of all Wick contractions of order $m$). For some order $m$, each contraction ($c$ denotes a specific contraction) in this sum results in a specific product of $m$ numbers of the form $\Theta^{XY}_{\alpha\beta}$ defined as
\begin{equation}
\begin{split}
        \Theta^{00}_{\alpha\beta} &= \sum_{\mu\in S^-}\chi^*_{\alpha,\mu}\chi_{\beta,\mu}\\
        \Theta^{01}_{\alpha\beta} &= \sum_{\mu\in S^-}\chi^*_{\alpha,\mu}\chi^*_{\beta,\mathcal{C}\mu}\\
        \Theta^{10}_{\alpha\beta} &= \sum_{\mu\in S^-}\chi_{\alpha,\mathcal{C}\mu}\chi_{\beta,\mu}=\Theta^{01*}_{\beta\alpha}\\
        \Theta^{11}_{\alpha\beta} &= \sum_{\mu\in S^-}\chi_{\alpha,\mathcal{C}\mu}\chi^*_{\beta,\mathcal{C}\mu} = \delta_{\alpha\beta}-\Theta^{00*}_{\alpha\beta}\\
\end{split}
\end{equation}
Each contraction $c$ of order $m$ corresponds to a permutation of the numbers $\{1,2,\cdots,2m\}$ under the following restriction: when the elements of the permutation are split in pairs $\{(a_k(c),b_k(c))\}_{k = 1}^{m}$ they must satisfy $a_k(c)< b_k(c)\in\{1,\dots,2m\}$ and $a_1(c)<a_2(c)<\cdots<a_m(c)$. Each pair yields $i_k(c) = \lfloor (a_k(c)+1)/2\rfloor$, $j_k(c) = \lfloor (b_k(c)+1)/2\rfloor$, $X_k(c) = (a_k(c)+1)\ \text{mod}\ 2$ and $Y_k(c) = b_k(c)\ \text{mod}\ 2$.  The overall sign $s(c)$ is the sign of the permutation. It is possible to write a script that procedurally generates all valid permutations and calculates the indices $X_k(c)$, $Y_k(c)$, $\alpha_{i_k(c)}$ and $\alpha_{j_k(c)}$ corresponding to every contraction $c$. 

\subsection{Convergence of parity}

The amount of terms in equation Eq.~\eqref{eq:parity_evolution} is
\begin{equation}
    1+\sum_{m=1}^{n_S}{n_S \choose m}(2m-1)!!.
\end{equation}
This number is out of reach in practice, so we are forced to truncate the sums. It was checked that restricting ourselves to order $m_{\text{max}}=4$ is sufficient to get an accurate result. In addition, the operator  $\langle P'\rangle$ defined in \ref{eq:bdg_parity_splitting} in principle contains $n_S=n-1$ Bogoliubov operators, but in practice we must truncate the product to a maximum number of states $n_\text{max}$, or equivalently, a cut-off energy $E_\mathrm{max}$. In Sec.~\ref{sec:parity}, we have argued that it is necessary to keep $E_\mathrm{max}<\Delta_\mathrm{J}$ so that $\langle\hat{P}'\rangle$ represents the parity of the edges. This is true for the case where $vt_\mathrm{inj}/W=2.3$ studied in Sec.~\ref{sec:charge}. We show this explicitly in Fig.~\ref{fig:parity_convergence}, where convergence is reached approximately at 0.85$\Delta_\mathrm{J}$, ensuring that no junction states participate in the calculation of the parity. We also show a few other cases with smaller values of $vt_\mathrm{inj}/W$. For these values, convergence of parity requires including up to 35 states with energies above $\Delta_\mathrm{J}$. In this case the calculation includes the hybridized edge and bound states of the junction, which does not allow us to isolate the edge parity sector from the junction.\\

\begin{figure}
    \centering
    \includegraphics[width=.5\linewidth]{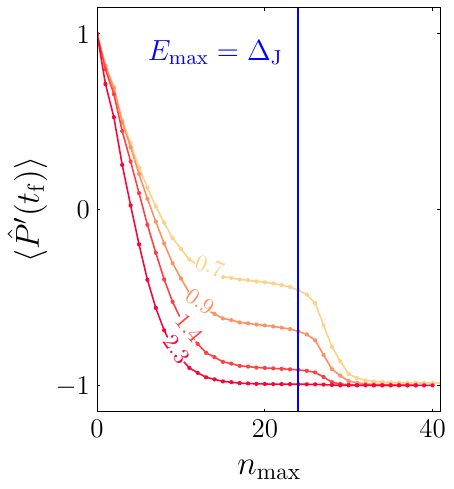}
    \caption{Convergence of the parity sector $\hat{P}'$ at final time $t_\mathrm{f}$ as a function of the index $n_\mathrm{max}$ which counts the number of eigenstates included in the calculation. This is done for different values of $vt_\mathrm{inj}/W$ which are displayed on the curves. For energies above $E_\mathrm{max}=\Delta_\mathrm{J}$, the hybridized edge-junction states are necessary in the convergence of the operator. }
    \label{fig:parity_convergence}
\end{figure}

\section{Supplemental results}
\label{app:exhaustive}
In this section, we present the results of our simulation for variable quenching times, supplementing the results in the main text. 

\subsection{Local representation of observables}
The calculations of current and quasi-particle number made in the main text have been integrated over specific areas. Here we show a few snapshots of the local current density and the local excitation density for two values of $vt_\mathrm{inj}/W$ (left and right panels of Fig.~\ref{fig:local_plots}). We show three different times in which the injection and fusion can be observed.

In the left panels, for long injections, the excitation entirely leaves the junction. In the right panel (which corresponds with Fig.~\ref{fig:excitation_density}), the excitation density slowly decays from the junction, at times even after $t>\tau=50a/v$ when the quench is over. 

The current density is zero in the superconducting region as the Majorana fermions are chargeless. Only upon fusion, the excitations produce charge. Here, the charge production at short injection times is much smaller, which is shown quantitatively in the next part. It is worth noting that while the excitations can remain trapped in the junction, they do not carry charge. 

\begin{figure*}
    \centering
    \includegraphics[width=\linewidth]{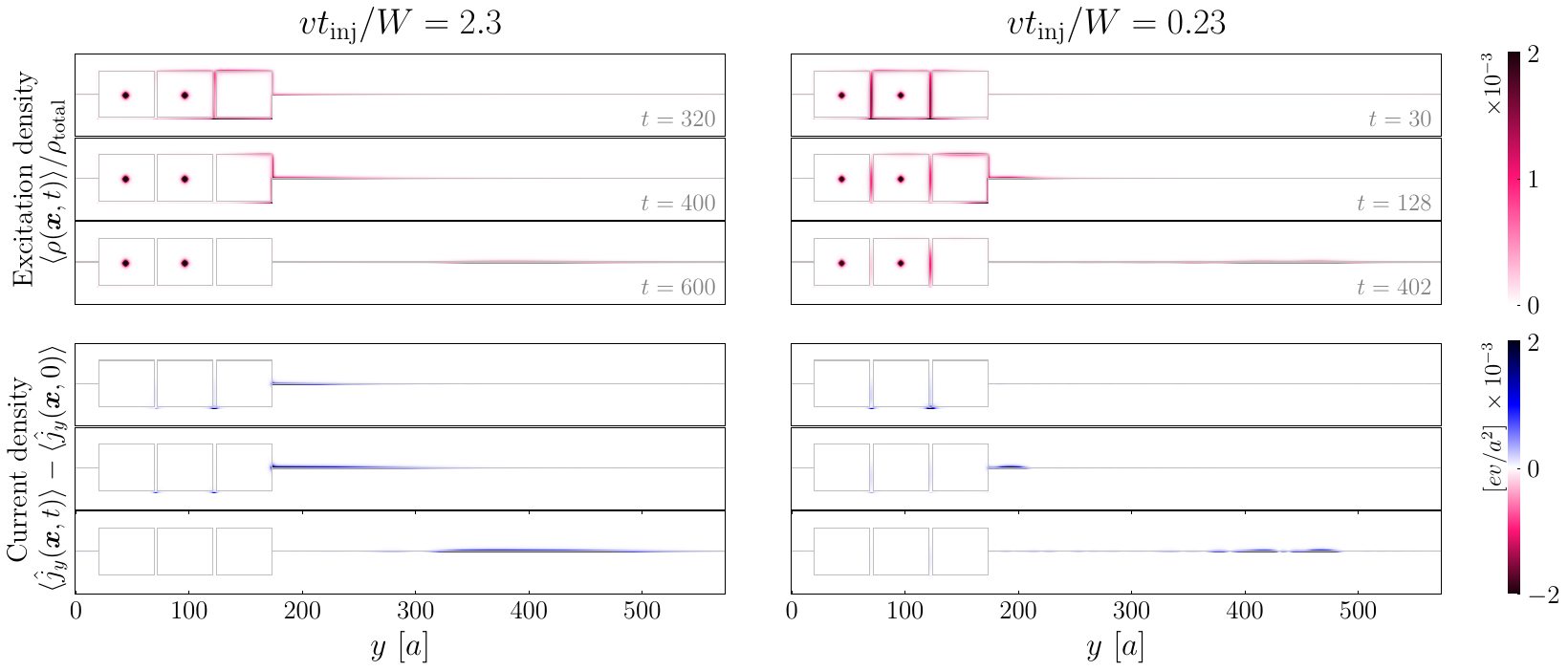}
    \caption{Three snapshots of the braiding protocol for two values of $vt_\mathrm{inj}/W$ (left column and right column). the top panel shows the snapshots in terms of the local excitation density, and the bottom panel shows them in terms of the local current density.}
    \label{fig:local_plots}
\end{figure*}

\subsection{Current density in the long junction regime}

For completeness, we include the calculations of charge at the exit for the different quenching times. In Fig.~\ref{fig:all_taus} we show the excitation spectrum, quasi-particle number, current and charge for different values of $vt_\mathrm{inj}/W$ discussed in Sec.~\ref{sec:bound_states}. We can see how the the occupancy of the junction increases when the injection time becomes shorter.

As the contribution of the excitations 
in the junction became sufficient the predictions for quantized charge are no longer valid. This can be seen in the bottom part of Fig.~\ref{fig:all_taus}  charge is no longer quantized. In the cases $vt_\mathrm{inj}/W=0.1,0.2$, not only the lowest mode but also the next higher mode of the junction is populated by excitations. Additionally as shown in Fig.~\ref{fig:local_plots}, a fast injection causes a large path-length difference as the junction traps the excitations and leaks them into the top and bottom edges at different rates. This results in further interference effects upon fusion.

\begin{figure*} 
    \centering
    \includegraphics[width=\linewidth]{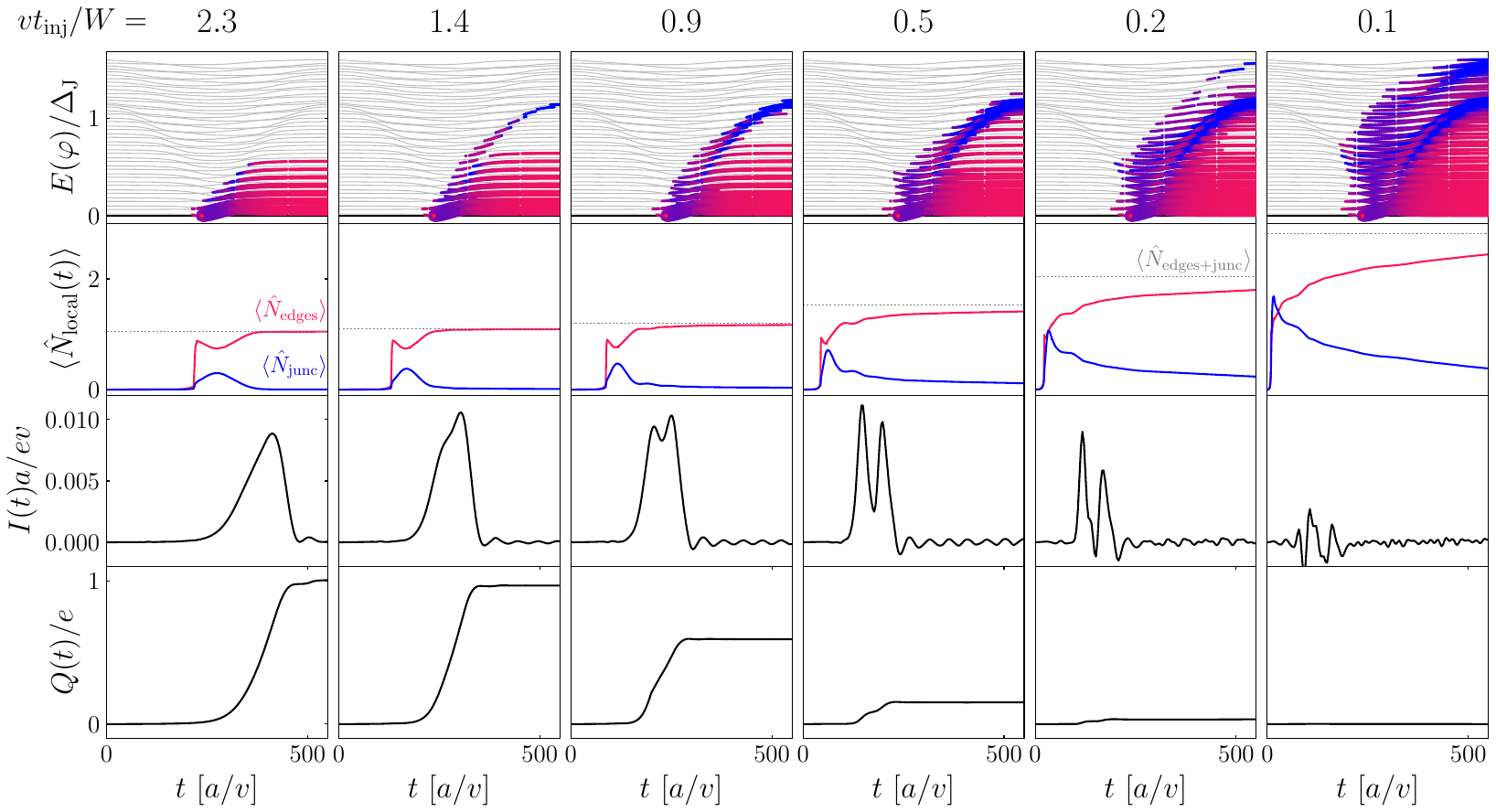}
    \caption{For different values of $vt_\mathrm{inj}/W$ from left to right, (First row) Quasi-particle occupation of the energy levels (thick colored lines). The color of the lines distinguishes between junction (blue) and edge (red) states.  (Second row) Quasi-particle number in the junction (blue), edges (red) and their sum at final value (gray dashed line). (Third row) Current at the exit of the superconductor. (Fourth row) Net charge creation at the exit of the superconductor. } 
    \label{fig:all_taus}
\end{figure*}

\end{document}